\documentclass[showpacs,preprintnumbers,amsmath,amssymb,prl,aps,twocolumn,superscriptaddress]{revtex4}

\usepackage{graphicx}
\usepackage{bm}
\usepackage[usenames,dvipsnames]{color} \usepackage{ulem}

\newcommand{\ligand}{{\underbar L}}

\begin{document}
\bibliographystyle{prsty}

\title{Charge disproportionation without charge transfer in the rare-earth 
nickelates as a possible mechanism for the metal-insulator transition}

\author{Steve Johnston}
\affiliation{Department of Physics and Astronomy, University of British Columbia, Vancouver, British Columbia, Canada V6T~1Z1}
\affiliation{Quantum Matter Institute, University of British Columbia, Vancouver, British Columbia, Canada V6T~1Z4}

\author{Anamitra Mukherjee}
\affiliation{Department of Physics and Astronomy, University of British Columbia, Vancouver, British Columbia, Canada V6T~1Z1}

\author{Ilya Elfimov}
\affiliation{Department of Physics and Astronomy, University of British Columbia, Vancouver, British Columbia, Canada V6T~1Z1}
\affiliation{Quantum Matter Institute, University of British Columbia, Vancouver, British Columbia, Canada V6T~1Z4}

\author{Mona Berciu}
\affiliation{Department of Physics and Astronomy, University of British Columbia, Vancouver, British Columbia, Canada V6T~1Z1}
\affiliation{Quantum Matter Institute, University of British Columbia,
  Vancouver, British Columbia, Canada V6T~1Z4}

\author{George A. Sawatzky}
\affiliation{Department of Physics and Astronomy, University of British Columbia, Vancouver, British Columbia, Canada V6T~1Z1}
\affiliation{Quantum Matter Institute, University of British Columbia,
  Vancouver, British Columbia, Canada V6T~1Z4}
\affiliation{Department of Chemistry, University of British Columbia, Vancouver, British Columbia, Canada V6T~1Z1}

\date{\today}

\begin{abstract}
We study a model for the metal-insulator (MI) transition in the
rare-earth nickelates RNiO$_3$, based upon a negative charge transfer
energy and coupling to a rock-salt like lattice distortion of the
NiO$_6$ octahedra. Using exact diagonalization and the Hartree-Fock
approximation we demonstrate that electrons couple strongly to these
distortions. For small distortions the system is metallic, with ground
state of predominantly $d^8\ligand$ character, 
where $\ligand$ denotes a ligand hole. For
sufficiently large distortions ($\delta d_{\rm Ni-O} \sim 0.05 -
0.10\AA$), however, a gap opens at the Fermi energy as the system
enters a 
periodically distorted state  alternating along the three
crystallographic axes, with $(d^8\ligand^2)_{S=0}(d^8)_{S=1}$
character, where  
$S$ is the total spin. Thus
the MI transition may be viewed as being driven by an  internal volume
``collapse'' where the NiO$_6$ octahedra with two ligand holes
shrink around their central Ni, while the remaining octahedra expand
accordingly, resulting in the ($1/2,1/2,1/2$) superstructure observed
in x-ray diffraction in the insulating phase. This insulating state is
an example of a new type of charge ordering achieved without any actual
movement of the charge. 
\end{abstract}

 \maketitle 


The perovskite rare-earth nickelates RNiO$_3$ undergo a first order
metal-insulator (MI) transition at a transition temperature
T$_{\mathrm{MI}}$ that can be tuned via pressure, strain, or
variations in the radius of the R ion (R $\ne$ La)
\cite{TorrancePRB1992, Review}.  These systems have received
considerable attention for over two decades as a means for studying MI
transitions; this has intensified in recent years due to proposals for
application in heterostructures
\cite{ChaloupkaPRL2008,HansmannPRL2009,LiuPRB2011}.  However, despite
this long history, the precise mechanism for the MI
transition is still not fully understood and a number of competing
proposals have been made.  These include a closing of the charge
transfer gap \cite{TorrancePRB1992}, Jahn-Teller polarons
\cite{MedardeIsotope}, the formation of long-range magnetic order (if
R = Nd, Pr) \cite{BalentsPRB2011,BalentsPRL2011}, charge 
ordering on the Ni sites, often referred to as charge
disproportionation (CD)
\cite{Review,AlonsoPRL1999,MedardePRB2008,MedardePRB2009,GarciaMunozPRB2009,
  AlonsoPRB2000,AlonsoPRB2001,AlonsoPRB2013}, or more complex charge
ordering distributed over the Ni and O sites, with the possible
involvement of the lattice
\cite{MizokawaPRB2000,ParkPRL2012,LauPRL2013}.

The leading interpretation for the MI transition is CD between the Ni sites. 
Several structural changes are concomitant 
with the MI transition, such as subtle changes in the Ni-O-Ni bond 
angle and changes in the Ni-O bond lengths resulting in a lowering of   
the crystal symmetry from an orthorhombic $Pbnm$ to a monoclinic $P2_1/n$ 
spacegroup \cite{Review,AlonsoPRB2000,AlonsoPRB2001,PiamontezePRB2005,
AlonsoPRB2013}. This forms two inequivalent Ni sites, where the 
NiO$_6$ octahedra contract and expand, respectively, about alternating
sites along  
the three crystallographic axes. Formal valence counting suggests that the 
Ni$^{3+}$ ions are  in a 3$d^7$ valence state and the inequivalent Ni sites 
are interpreted as a 3$d^{7}$3$d^{7}$ $\rightarrow$ 
3$d^{(7-\delta)}$3$d^{(7+\delta)}$ CD below T$_{\mathrm{MI}}$ 
\cite{MedardePRB2008, MedardePRB2009, GarciaMunozPRB2009, AlonsoPRB2013}. 
This CD is thought to drive the MI transition, however 
it is not clear whether the electron-lattice or the on-site electron-electron 
interactions drive the charge ordering.  
T$_\mathrm{MI}$ exhibits a large $^{16}$O - $^{18}$O isotope effect 
\cite{MedardeIsotope}, confirming that the lattice 
and electronic degrees of freedom are strongly coupled, 
and the latter could be driven by the former.
A recent proposal views the lattice distortion as a secondary effect to a   
magnetic ordering \cite{BalentsPRB2011,BalentsPRL2011}, however 
this picture is difficult to apply to systems with  
T$_\mathrm{N}$ $<$ T$_\mathrm{MI}$, which is the case across much of the 
phase diagram \cite{Review}.

The CD scenario assumes that Ni 3$d^{7}$ is the appropriate
starting point for describing the nickelates, but a few open issues
remain.  In the (nearly) cubic structure, the Ni 3$d^{7}$
($t_{2g}^6e_g^1$) ion has two degenerate $e_g$ orbitals which the
system will prefer to lift via a coherent Jahn-Teller effect or via CD
between neighbouring Ni sites.  There is no sign 
of a JT distortion \cite{MedardeIsotope}
like the orbital ordering in manganites
\cite{Maganites}, which also have a singly occupied degenerate $e_g$
orbital. On the other hand, CD should be suppressed by the large
Hubbard $U$, however it can be stabilized by a sizeable Hunds coupling
if the system is close to the itinerant limit
\cite{MazinPRL2007}. This  is likely the case in the
rare-earth nickelates, which have been classified as small or negative
charge transfer systems
\cite{MizokawaPRB1995,MizokawaPRB2000,ParkPRL2012,LauPRL2013,Bodenthin2011,
  AbbatePRB2002,HoribePRB2007,HanPRB2012,BlancaRomeroPRB2011,AnisiovPRB1999}.
In such conditions, however, it becomes energetically favourable to
transfer a hole from Ni to ligand O sites \cite{ZaanenPRL1985},
and Ni $3d^7$  is no longer the appropriate starting point.
Rather, one should begin with a Ni $d^8\ligand$ state in the atomic
limit, which is equivalent to taking a negative charge transfer energy
before allowing for hybridization between the Ni and O atoms. In this
picture, the ground state would have, on average, one ligand
hole per NiO$_3$ unit.

Ligand holes can couple strongly to the bond-stretching O
vibrations which modulate the hybridization between the 
TM and  O sites
\cite{MedardeIsotope,Gunnarsson}. Thus, the carriers  in
 nickelates (which are primarily O ligand holes in our picture)  
can couple strongly to the rock-salt-like lattice distortions. In this
Letter we examine this  
scenario as a means for explaining the MI transition. 
Using exact diagonalization (ED) and Hartree-Fock  (HF)
calculations \cite{Chen, MeanField} we   
confirm that a strong electron-lattice coupling drives the system 
through a MI transition for sufficiently large and experimentally 
relevant lattice distortions 
($\delta d_{\rm Ni-O} \sim 0.05$-$0.1\AA$). In the insulating phase 
we find a novel charge ordering 
which can be viewed as a $(d^8\ligand)_i(d^8\ligand)_j$ $\rightarrow$  
$(d^8\ligand^2)_{S=0}(d^8)_{S=1}$ type, where $S$ is the total spin.  
In this picture, the MI transition results from a
partial volume collapse of NiO$_6$ octahedra with two ligand holes 
around their central Ni, while the other octahedra
expand accordingly with little net effect on the total volume,
as observed experimentally \cite{Review}. However, since each O is shared by a
collapsed and an expanded octahedron, in fact all octahedra
have the same average hole
concentration. The difference is that the two ligand
holes acquire the symmetry of the $e_g$ orbitals of the Ni in the
collapsed $d^8\ligand^2$ octahedron, which therefore becomes
formally equivalent to a 3$d^6$ state that is effectively no longer 
JT active, and attains a spin of zero.    
Thus, as far as symmetry and spin are concerned, this new charge state
appears as a fully CD state, but one 
achieved without actual movement of charge.
We note that a similar charge/spin distribution was proposed 
in Refs. \onlinecite{ParkPRL2012,LauPRL2013}, but as being driven by very
different mechanisms.

\begin{figure}
 \includegraphics[width=\columnwidth]{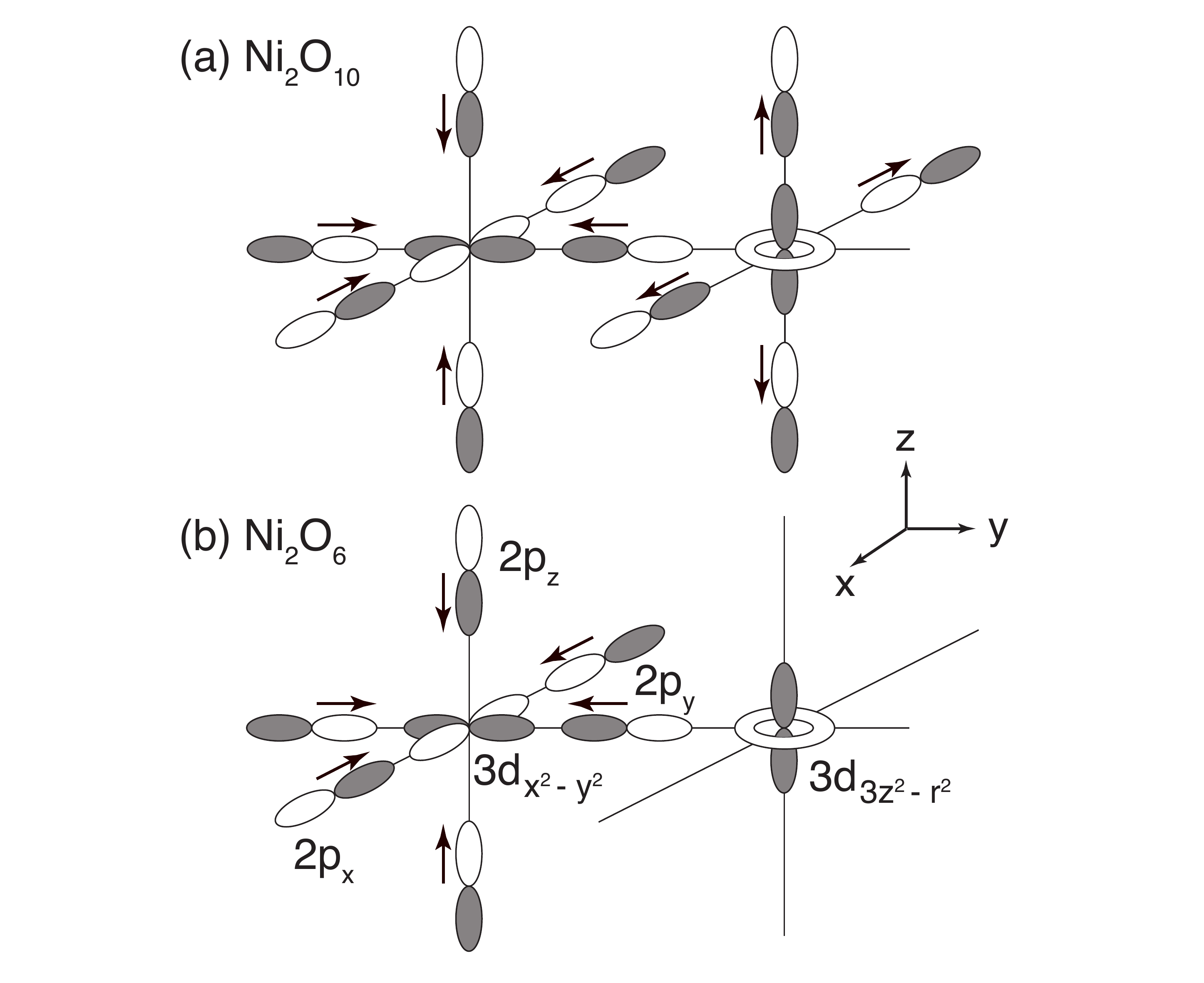}
 \caption{\label{Fig:clusters}  
  (a) 
 Ni$_2$O$_{10}$ quasi-1D cluster with periodic boundary conditions
 along the $y$-axis, studied with ED. Both Ni 3$d$ ${\rm e}_{\rm g}$
 orbitals are included at each Ni site, but only one orbital per Ni is
 depicted here to establish our phase convention. (b)
 Ni$_2$O$_6$ unit cell used for our  3D calculations using HFA.
 The arrows in both panels indicate the displacement pattern for the
 rock-salt-like oxygen distortion.
 }
\end{figure}

We investigate quasi-one-dimensional (1D) 
and three-dimensional (3D) clusters/unit cells.
In each case, the orbital basis includes  
both 3$d$ $e_g$ orbitals at each Ni site and the 2p$_\sigma$ 
orbital at each O site, as shown in Fig. \ref{Fig:clusters}. 
The microscopic Hamiltonian for the electronic degrees of freedom 
is $H = H_{\rm 0} + H_{\rm int}$, where 
\begin{equation*}
    H_{\mathrm 0} = \sum_{i,\alpha} \epsilon_\alpha 
    d^\dagger_{i,\alpha,\sigma}d^{\phantom\dagger}_{i,\alpha,\sigma}
    +\sum_{i,j,\alpha,\beta,\sigma} t^{\alpha\beta}_{ij} 
    d^\dagger_{i,\alpha,\sigma}d^{\phantom\dagger}_{j,\beta,\sigma}
\end{equation*}
contains the non-interacting on-site and near-neighbor hopping terms, and 
\begin{eqnarray*}
H_{\rm {int}}&=&\sum_{i,\alpha,\sigma\ne\sigma^\prime} \frac{U}{2} n_{i,\alpha,\sigma}n_{i,\alpha,\sigma^\prime} 
+ \sum_{i,\alpha\ne\alpha^\prime,\sigma,\sigma^\prime}\frac{U^\prime}{2} 
 n_{i,\alpha,\sigma}n_{i,\alpha^\prime,\sigma^\prime} \\
&+&\sum_{i,\alpha,\alpha^\prime,\sigma,\sigma^\prime} \frac{J}{2}
d^\dagger_{i,\alpha,\sigma}d^\dagger_{i,\alpha^\prime,\sigma^\prime}
d^{\phantom\dagger}_{i,\alpha,\sigma^\prime}d^{\phantom\dagger}_{i,\alpha^\prime,\sigma}\\
&+&\sum_{i,\alpha\ne\alpha^\prime,\sigma\ne\sigma^\prime} \frac{J^\prime}{2}
d^\dagger_{i,\alpha,\sigma}d^\dagger_{i,\alpha,\sigma^\prime}
d^{\phantom\dagger}_{i,\alpha^\prime,\sigma^\prime}d^{\phantom\dagger}_{i,\alpha^\prime,\sigma}
\end{eqnarray*}
contains the on-site Coulomb interactions with 
$n_{i,\alpha,\sigma} =
d^\dagger_{i,\alpha,\sigma}d^{\phantom\dagger}_{i,\alpha,\sigma}$,
where $ d^\dagger_{i,\alpha,\sigma}$ creates a spin-$\sigma$ electron
in one of the two $e_g$ orbitals if $i$ is a Ni site, or the
$2p_\sigma$ orbital if $i$ is an O site. In eV units we set 
$U_d = 7$, $J_d = J^\prime_d = 0.9$, $U_d^\prime = U_d - 2J_d$, 
$U_p = 5$, $\epsilon_d = 0$, $\epsilon_p = 2.25$, $(pd\sigma) = -1.8$, 
$(pp\sigma) = 0.6$, $(pd\pi) = -(pd\sigma)/2$, and $(pp\pi) = -(pp\sigma)/4$. 
For these parameters the  ground state of
an isolated NiO$_6$ octahedron is
$|\Psi\rangle = \alpha |d^7\rangle
+\beta|d^8\ligand\rangle + \gamma|d^9\ligand^2\rangle$ with 
$\beta^2 \approx 0.56$, i.e. the system is in the negative 
charge transfer regime and the ground state has
predominantly $d^8\ligand$  character, in agreement with 
experiment \cite{MizokawaPRB1995,AbbatePRB2002,HoribePRB2007,Bodenthin2011}.  

The lattice degrees of freedom are treated in the adiabatic limit 
as frozen lattice displacements \cite{phonons}, i.e. all O atoms are
displaced equally  
along the rock-salk displacement pattern 
indicated by arrows in Fig. \ref{Fig:clusters}. The electron-lattice 
coupling is introduced by rescaling the Ni-O and O-O 
transfer integrals. Specifically, we take 
$t_{pd} = t_{pd}^0(1 + \delta d_{\rm Ni-O}/d_{\rm Ni-O})^{-3}$ and 
$t_{pp} = t_{pp}^0(1 + \delta d_{\rm O-O}/d_{\rm O-O})^{-4}$
\cite{Harrison, LauPRL2013},  
where $\delta d$ is the change in the appropriate bond length 
with respect to $d_{\rm Ni-O} = d_{\rm O-O}/\sqrt{2} = 1.95$ $\AA$ in the 
undistorted structure. The lattice potential energy adds an overall 
quadratic term to the total energy, while the lattice kinetic energy
is neglected.    
From now on all displacements are reported in terms of  $\delta
d_{\rm Ni-O}$, and we drop its subscript for 
brevity. We note that we have also tested the other possible displacement 
patterns for the O atoms, however the static rock-salt-like pattern discussed
here produces the lowest energy \cite{Supplement}.

\begin{figure}
 \includegraphics[width=0.8\columnwidth]{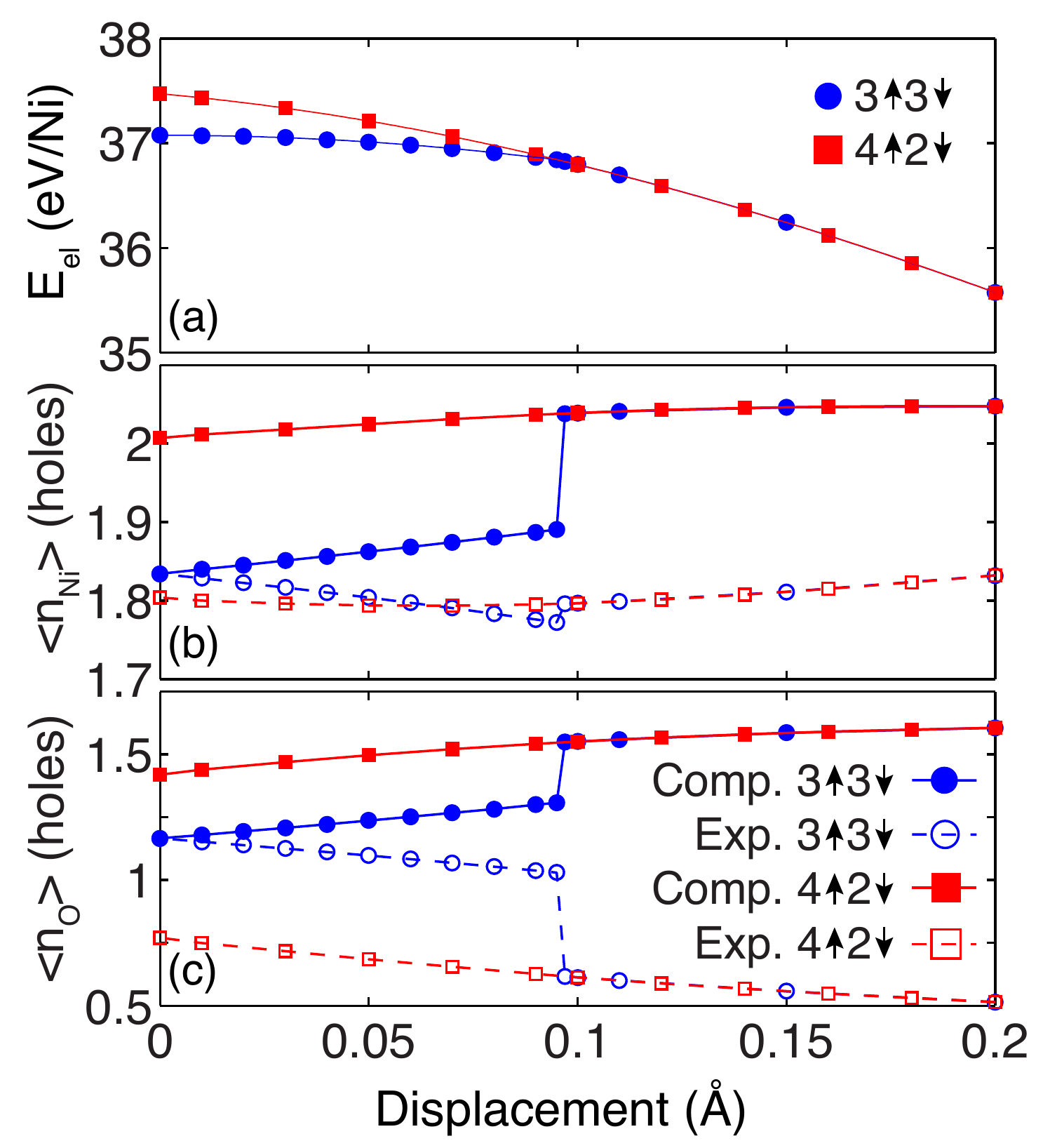}
 \caption{\label{Fig:ED_Chain} (color online)
 A summary of ED results for the Ni$_2$O$_{10}$ cluster. 
 (a) Electronic contribution to the ground state energy in the
 ($3\uparrow 3\downarrow$) (blue $\bigcirc$)  
 and ($4\uparrow2\downarrow$) (red $\square$) sectors as a function of the 
 static O displacement. (b), (c) Average hole occupancy at Ni 
 and O sites, respectively.  Sites on the left (right)  
 side of the Ni$_2$O$_{10}$ cluster correspond to the compressed 
 (expanded) NiO$_6$ 
 octahedron for $\delta d >0$ (see Fig. \ref{Fig:clusters}), 
 and are represented by the solid (open) symbols.  
 }   
\end{figure}

We first consider two neighbouring NiO$_6$ octahedra 
in the quasi-1D cluster of Fig. \ref{Fig:clusters}(a) with three holes/Ni;  
this can be solved exactly using ED.  
Fig. \ref{Fig:ED_Chain}(a) shows the electronic 
contribution to its ground state energy, $E_{el}$,
in the $(3\uparrow3\downarrow)$ and $(4\uparrow2\downarrow)$  
sectors as a function of $\delta d$.  Figs. \ref{Fig:ED_Chain}(b), (c)
show average hole  occupations at Ni and O sites   
for the compressed (solid symbols) and expanded (open symbols) NiO$_6$ octahedra. 
Both sectors exhibit a sizable decrease in the electronic energy with increasing    
displacement, indicating a rather strong electron-lattice coupling. 
The $(3\uparrow3\downarrow)$ sector exhibits a kink in $E_{el}$ 
for a critical displacement $\delta d_c\sim 0.097$ $\AA$, indicative
of a level crossing and a change in the ground state  
wave function as a function of displacement. 

For $\delta d = 0$ the ground state has a 
total spin $S_{\rm tot} = 0$ with majority $(d^8\ligand)(d^8\ligand)$ 
character, as expected for a negative value of $\Delta$. This 
state persists as the ground state for small $\delta d < \delta d_c$,
however a small  
redistribution of charge occurs between the two octahedra due to 
the increased delocalization on the compressed side.   
For $\delta d > \delta d_c$ a different ground state is obtained with 
$S_{\rm tot} = 1$ and $m_z = 0$.  
(In the $4\uparrow2\downarrow$ sector the $S_{\rm tot} = 1$, $m_z = \pm 1$ 
counterparts of this state form the degenerate ground state for 
all values of $\delta d$.) 
The new state possesses an interesting redistribution of charge. 
Both Ni ions are in a $d^8$ spin triplet state while 
the $\ligand$ hole of the expanded octahedron transfers to the compressed side, 
where the two $\ligand^2$ holes 
lock in a triplet pair due to the finite value of $U_{p}$ \cite{ElfimovPRL2002}. 
The $\ligand^2$ spin couples antiferromagnetically to the central Ni
spin, leaving the  
system in a $(d^8\ligand^2)_{S = 0}(d^8)_{S=1}$ state.  
Thus, for sufficiently large rock-salt-like displacements, 
the ligand hole of the expanded octahedron can re-associate with a neighbouring 
NiO$_6$ octahedron leaving a lone $S = 1$ spin on alternating Ni sites  
and a small CD $\sim 0.1$ between the two inequivalent Ni sites. 
We note that the critical value $\delta d_c$ obtained here is slightly larger 
than the   
typical changes in bond length $\sim 0.052$ - $0.092$ $\AA$ reported 
in the insulating phase \cite{MedardePRB2008,GarciaMunozPRB2009, AlonsoPRB2013}. 

Below the MI transition a rock-salt distortion exists in the
nickelates and  a 
similar re-association of the ligand holes may occur in the 3D
system as well. However, while for the Ni$_2$O$_{10}$ cluster we can assign 
the O sites  to unique octahedra, in the bulk 3D system this
is not possible since each O is shared by an expanded and a contracted
octahedron. The $\ligand^2$ triplet pairs assigned to the collapsed
NiO$_6$ octahedra are formed between the  ligand hole of the compressed
NiO$_3$ unit and a second hole provided by the neighbouring
expanded NiO$_3$ units. From a charge counting perspective, each of
the NiO$_3$ units still has one ligand hole and hence the system
appears to have dominant $d^8\ligand$ character. What is changed
is not the ligand holes' average distribution, but the phases in their wavefunction
which now acquire the symmetry of the two $e_g$ orbitals of the Ni
in the collapsed octahedron. The basic effect is similar to the Zhang-Rice 
singlet scenario in the cuprates where the ligand hole selects the 
relative phases of the four oxygen orbitals so as to maximally 
hybridize with one of the two Cu ions bridged by the oxygen. 

We stress that
this is a very different form of charge ordering than the commonly
discussed CD; it should be thought of as a novel charge ordering
without a significant movement of charge.

\begin{figure*}[ht]
 \includegraphics[width=\textwidth]{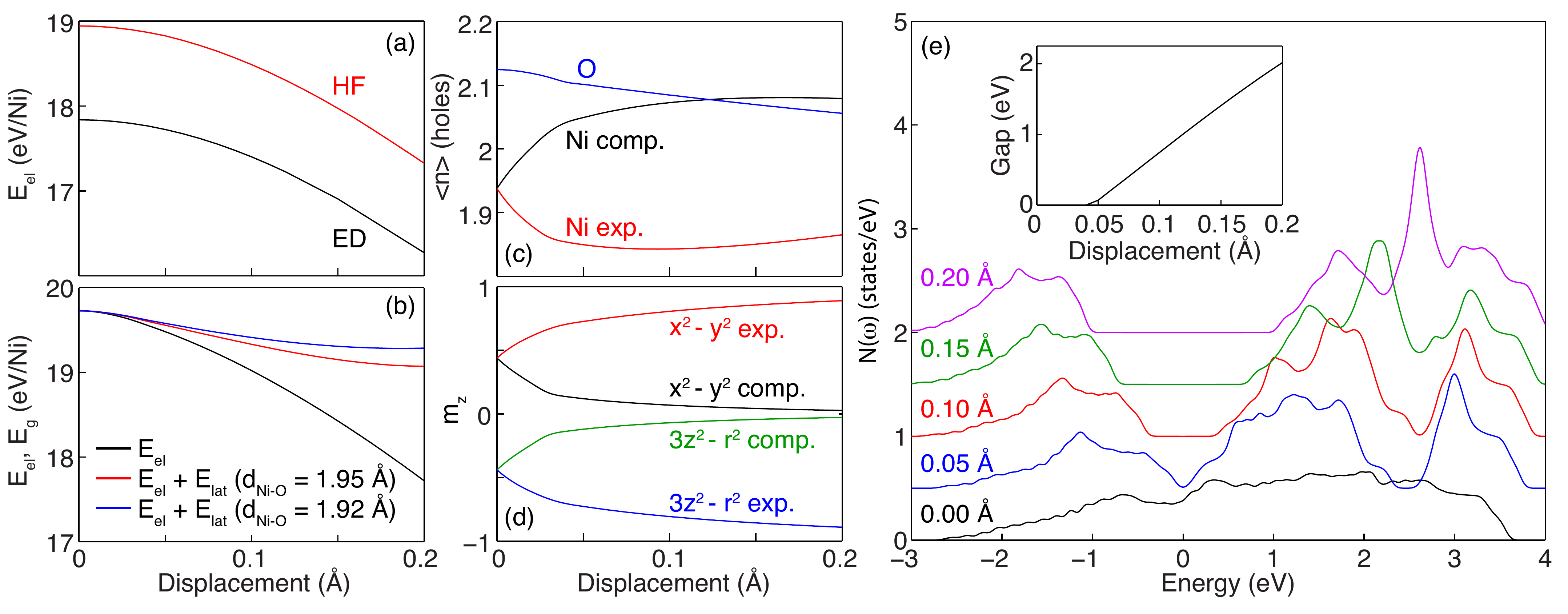}
 \caption{\label{Fig:DOS} (color online) (a) A comparison of the
   dependence of the HFA (red/light) and ED (black/dark) ground state
   energies of the Ni$_2$O$_6$ cluster as a function of the oxygen
   displacement.  (b)-(e) HFA results for the 3D lattice with
   $80\times80\times80/2$ momentum points in the Brillouin zone:
   (b) ground state electronic energy, (c) average hole occupation of the oxygen
   (blue) sites and the compressed/expanded Ni sites (black/red), (d)
   magnetization $m_z = (n_\uparrow-n_\downarrow)$ of the Ni orbitals,
   (e) single particle density of states for several oxygen
   displacements.  The spectrum has been lifetime broadened with a
   Lorentzian (HFHM $=$ 25 meV). Another band of states is located
   below 3 eV (not shown). The inset shows the gap as a function of
   displacement. 
}
\end{figure*}

We now turn to a  quantitative analysis of the 3D case, and
examine an extended lattice that cannot be handled using 
ED. We therefore implement an unbiased HFA \cite{MeanField}
(see the supplementary material for details
\cite{Supplement}). One might doubt the validity of the HFA
given the strong correlations on the Ni sites, however the large bandwidth O 2p
bands overlap with the Ni states
because of the negative charge transfer energy. This reduces significantly the
effects of correlations, explaining why a number of studies have had
success with mean-field approaches
\cite{MizokawaPRB1995,MizokawaPRB2000,LauPRL2013}.  To verify that the HFA  
captures the physics of our model we compare its predictions to ED results for
a Ni$_2$O$_6$ cluster with periodic boundary
conditions, see Fig.  \ref{Fig:clusters}(b). $E_{el}$ results are shown in
Fig. \ref{Fig:DOS}(a). Not surprisingly, the HFA has higher 
ground state energies ($\sim 1$ eV), but it correctly reproduces the
displacement dependence and the average hole occupations 
\cite{Supplement}. This gives us confidence that HFA 
is reasonably accurate for this Hamiltonian.

Scaling to the extended lattice, Figs. \ref{Fig:DOS}(b)-(e) show HF results
obtained for a 3D system with $0.5\times40^3$ momentum points in the
 Brillouin zone. The electronic component of the ground state energy,
 $E_{el}$, 
(Fig. \ref{Fig:DOS}b) essentially mirrors the ED results, exhibiting
large variation with lattice displacement.  This confirms the strong
electron-lattice coupling  even within the HFA.  The hole
occupancies $n$ (Fig. \ref{Fig:DOS}c) and Ni magnetization $m_z$
(Fig. \ref{Fig:DOS}d) also follow the expectations gained from the
Ni$_2$O$_{10}$ cluster.  For small $\delta d$ the total
occupancies of the two Ni sites start to separate until $\delta d \sim
0.03$ - $\sim 0.04$ $\AA$, when they gradually level off. At
the same time the magnetizations of the two $e_g$ orbitals in the
compressed octahedron fall to zero as each $e_g$ orbital is occupied
by half an electron of each spin species, on average. Similarly, the
magnetization of each $e_g$ orbital in the expanded octahedron rapidly
grows to nearly one but with opposite alignments. The magnetization of
the O sites (not shown) is nearly zero for all values of $\delta
d$.  Although the HFA cannot reproduce the triplet
correlations of the exact cluster solution, these results mimick
the $(d^8\ligand^2)_{S = 0}(d^8)_{S=1}$ ground state observed in 
the small cluster and therefore confirm our picture
of the novel charge ordering.

Next, we calculate the single particle density of states $N(\omega)$
as a function of $\delta d$, see Fig. \ref{Fig:DOS}(e). For $\delta
d=0$, $N(\omega)$ has finite weight at the Fermi energy $\omega=0$,
consistent with metallic behavior in a negative charge transfer
system. $N(\omega=0)$ is suppressed  for increasing $\delta d$ until
$\delta d \approx 0.05$ $\AA$, where a gap that increases linearly
with $\delta d$ opens up (inset, Fig. \ref{Fig:DOS}(e)).  We stress
that this gap is the result of the alternating expanded and
contracted octahedra rather than the formation of the $(S =
0)_i(S = 1)_j$ magnetic structure, because similar sized gaps are
observed even when we restrict the HFA by enforcing identical Ni sites. 
We conclude that the gap opens due
to the electron-lattice interaction and rock-salt lattice distortions,
consistent with the MI transition observed in these materials.

We complete our discussion with an estimate for the energy cost of the
octahedral distortions, based on density functional theory (DFT)
within the generalized gradient approximation (GGA)\cite{gga_PBE,wien2k}. The
total energy was calculated for the cubic LaNiO$_3$ structure with
lattice constants of 3.9 $\AA$ and 3.84 $\AA$, the latter being the
DFT prediction \cite{DFT_details}. In both cases we find the change in
total energy to increase like $(\delta d)^2$.  We use this change in
total energy as a very rough estimate of the potential energy
of the lattice distortion \cite{Footnote}.
Fig. \ref{Fig:DOS}(b) shows the total energy obtained by adding this
contribution to $E_{el}$ obtained from the HFA. The total energy
possesses a deep minimum for $\delta d \sim 0.17-0.2$ $\AA$.  This is
a factor of two larger than the experimental value, but reasonable
given the crudeness of our estimate for the lattice contribution. 
Note that a strong isotope effect is obviously consistent with 
this model of the MI transition.


In summary, we have examined the consequences of the interplay between
a negative charge transfer energy and the electron-lattice coupling in
RNiO$_3$, starting from a $d^8\ligand$ atomic configuration.  Our results
demonstrate that in this 
parameter regime a strong electron-lattice interaction arises and that
it drives a MI transition.  In this picture, the nickelates 
undergo a new type of effective CD achieved without any significant movement
of charge between neighbouring NiO$_6$ octahedra.  In order to verify
whether such a picture extends across the whole family of nickelates,
systematic studies capable of incorporating  more
accurately the lattice potential are required.

{\it Acknowledgements - } 
We thank M. W. Haverkort for useful discussions. This work was 
supported by NSERC, CIfAR, and the Max Planck - UBC Centre for Quantum Materials.


\begin{thebibliography}{99}
\bibitem{phonons} This can also be viewed as a variational calculation
  where the quantum phonons at each O site are in the coherent state
  consistent with its displacement.
\bibitem{DFT_details} DFT calculations were performed using code
WIEN2K\cite{wien2k} and FCC ($Fm\bar{3}m$) supercell with 2 formula units.
Self-consistency was obtained with 1000 k points in the reduced Brillouin
zone.
\bibitem{Footnote}
This should be regarded as a very rough estimate as the change in covalency 
due to the lattice deformation is doubly counted in this approach.
\end{thebibliography}
\end{document}


\bibliographystyle{prsty}

\title{Charge disproportionation without charge transfer in the rare-earth 
nickelates as a possible mechanism for the metal-insulator transition - 
supplementary material}

\author{Steve Johnston}
\affiliation{Department of Physics and Astronomy, University of British Columbia, Vancouver, British Columbia, Canada V6T~1Z1}
\affiliation{Quantum Matter Institute, University of British Columbia, Vancouver, British Columbia, Canada V6T~1Z4}

\author{Anamitra Mukherjee}
\affiliation{Department of Physics and Astronomy, University of British Columbia, Vancouver, British Columbia, Canada V6T~1Z1}

\author{Ilya Elfimov}
\affiliation{Department of Physics and Astronomy, University of British Columbia, Vancouver, British Columbia, Canada V6T~1Z1}
\affiliation{Quantum Matter Institute, University of British Columbia, Vancouver, British Columbia, Canada V6T~1Z4}

\author{Mona Berciu}
\affiliation{Department of Physics and Astronomy, University of British Columbia, Vancouver, British Columbia, Canada V6T~1Z1}
\affiliation{Quantum Matter Institute, University of British Columbia,
Vancouver, British Columbia, Canada V6T~1Z4}

\author{George A. Sawatzky}
\affiliation{Department of Physics and Astronomy, University of British Columbia, Vancouver, British Columbia, Canada V6T~1Z1}
\affiliation{Quantum Matter Institute, University of British Columbia,
Vancouver, British Columbia, Canada V6T~1Z4}
\affiliation{Department of Chemistry, University of British Columbia, Vancouver, British Columbia, Canada V6T~1Z1}

\date{\today}

\pacs{71.30.+h, 71.38.-k, 72.80.Ga} \maketitle 

\section{Phonon displacement pattern}
In order to verify that the rock-salt-like distortion pattern assumed
in the main text is energetically favored we performed a series of ED
calculations for the Ni$_2$O$_{10}$ cluster. In each case we allowed
the ten O atoms to be displaced by $\delta d = \pm 0.1$
$\AA$ along the Ni-O bond direction. ED calculations for the ground
state energy were then carried out for the $2^{10}$ possible
configurations. The lowest energies were obtained for the two
displacement patterns corresponding to a contraction/expansion or
expansion/contraction of the O octahedra about the two Ni sites as 
considered in the article.

\section{Hartree-Fock Calculations}
We performed unbiased Hartree-Fock calculations for the model
Hamiltonian given in the main text.   
Our treatment follows that of Ref. \onlinecite{MeanField}, which we 
summarize here for completeness. We work in momentum space where the 
microscopic Hamiltonian can be rewritten in the form 
\begin{equation}
H=\sum_{\bk,\alpha,\beta,\sigma} T_{\alpha,\beta}(\bk) d^\dagger_{\bk,\alpha,\sigma}
d^{\phantom\dagger}_{\bk,\beta,\sigma} 
+ \sum_{\alpha,\alpha^\prime,\beta,\beta^\prime,\sigma,\sigma^\prime}\sum_{\bk,\bk^\prime,\bq} 
U^{\sigma,\sigma^\prime}_{\alpha,\alpha^\prime,\beta,\beta^\prime}(\bq)  
d^{\dagger}_{\bk,\alpha,\sigma}d^{\dagger}_{\bk^\prime,\alpha^\prime,\sigma^\prime}
d^{\phantom\dagger}_{\bk^\prime-\bq,\beta^\prime,\sigma^\prime}
d^{\phantom\dagger}_{\bk+\bq,\beta,\sigma}. 
\end{equation}
where the interaction on a Ni site is given by  
\begin{eqnarray}
U^{\sigma\sigma^\prime}_{\alpha,\alpha^\prime,\beta,\beta^\prime}&=& 
 \frac{U}{2} \delta_{-\sigma,\sigma^\prime}\delta_{\alpha,\alpha^\prime}\delta_{\alpha\beta}\delta_{\alpha\beta^\prime}  
 +\frac{U^\prime}{2}(1-\delta_{\alpha\alpha^\prime}) \delta_{\alpha\beta}\delta_{\alpha^\prime\beta^\prime} \\\nonumber
 &+&\frac{J}{2}(1-\delta_{\alpha\alpha^\prime})\delta_{\alpha\beta^\prime}\delta_{\alpha^\prime\beta} + 
 \frac{J^\prime}{2}\delta_{\alpha\alpha^\prime}\delta_{\beta\beta^\prime}(1-\delta_{\sigma\sigma^\prime})(1-\delta_{\alpha\beta}).
\end{eqnarray} 
The interaction on an O site is like the first term with $U$ replaced by $U_p$. 

Our aim is to now construct a HF Hamiltonian of the form  
\begin{equation}
    H_{\rm HF} = \sum_{n,\bk} E_n(\bk)\gamma^\dagger_{n,\bk,\sigma}\gamma^{\phantom\dagger}_{n,\bk,\sigma}, 
\end{equation}
where $n$ is a band index and $\bk$ is the electron momentum
restricted to the first Brillouin zone. 
We begin by expanding the HF operators in terms of the original basis set 
$\gamma_{\bk,n,\sigma} = \sum_{\alpha}
c_{n,\alpha}(\bk)d_{\bk,\alpha,\sigma}$, and then obtain the HF matrix
equation for the expansion coefficients   
$\sum_{\alpha} M_{\alpha\beta}(\bk) c_{\alpha}(\bk) = E_n(\bk)c_\alpha(\bk)$, 
which is solved self-consistently. The matrix $M_{\alpha\beta}(\bk)$
results from the equation:
\begin{equation}
   \langle \big\{[\gamma^{\phantom\dagger}_{n,\bk,\sigma},H], d^\dagger_{\bk,\alpha^\prime,\sigma^\prime}\big\} \rangle= 
    \big\{ E_n(\bk)\gamma^{\phantom\dagger}_{n,\bk,\sigma},d^\dagger_{\bk,\alpha^\prime,\sigma^\prime} \big\}.
\end{equation} 
For a quadratic Hamiltonian the left-hand side result of the
commutators/anticommutators is a c-number and no average would be
needed. For a quartic Hamiltonian that side contains operators which
are replaced with their HF values. This method leads to  the true
unbiased HF solution
(i.e., the Slater determinant that minimizes the total energy, see
\cite{MeanField}). 

After lengthy but straightforward algebra we find:
\begin{eqnarray}
M_{\alpha\beta}(\bk)=T_{\alpha,\beta}(\bk)&+&\frac{1}{2}\sum_{\bq,\alpha^\prime,\beta^\prime,\sigma^\prime} 
    \left[ U^{\sigma\sigma^\prime}_{\beta,\alpha^\prime,\alpha,\beta^\prime} + 
           U^{\sigma^\prime\sigma}_{\alpha^\prime,\beta,\beta^\prime,\alpha}
    \right] \langle d^\dagger_{\bq,\alpha^\prime,\sigma^\prime}d^{\phantom\dagger}_{\bq,\beta^\prime,\sigma^\prime} \rangle 
    \\ \nonumber 
 &-&\frac{1}{2}\sum_{\bq,\alpha^\prime,\beta^\prime} 
    \left[U^{\sigma\sigma}_{\alpha^\prime,\beta,\alpha,\beta^\prime} + 
          U^{\sigma\sigma}_{\beta,\alpha^\prime,\beta^\prime,\alpha}\right]
    \langle d^\dagger_{\bq,\alpha^\prime,\sigma}d^{\phantom\dagger}_{\bq,\beta^\prime,\sigma} \rangle 
\end{eqnarray}
where the average $\langle  \rangle$ defines the
self-consistent HF fields, which  are found iterationally. 
 It should be noted that the dependence on the static lattice displacement is 
introduced through the kinetic term $T_{\alpha,\beta}(\bk)$, see main text.

\section{Comparison with Exact Diagonalization}
\begin{figure}
 \includegraphics[width=0.5\columnwidth]{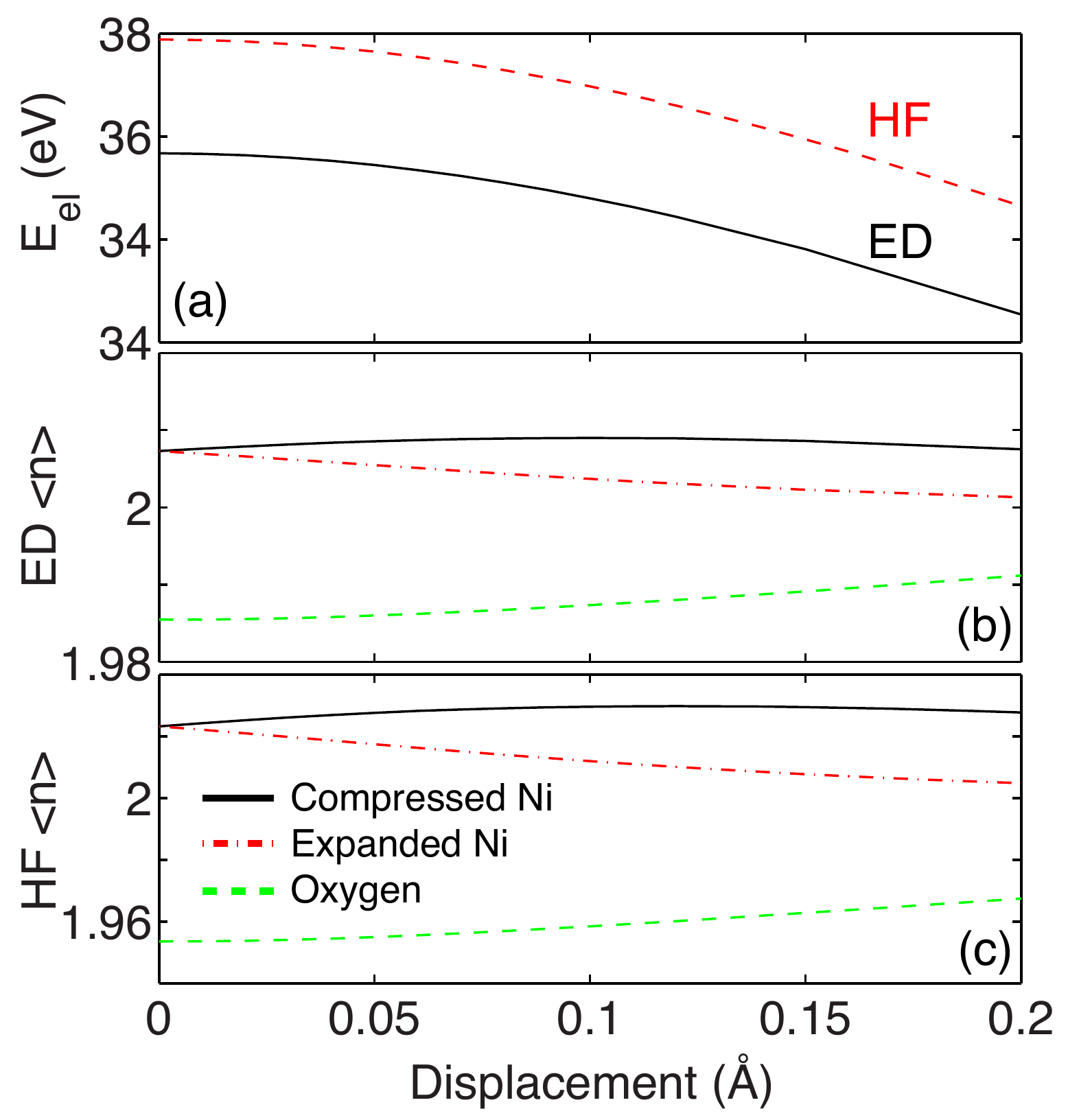}
 \caption{\label{Fig:MF_vs_ED} (color online)
 (a) The ground state energy per Ni atom of the 
 Ni$_2$O$_6$ cluster with periodic boundary conditions. 
 HF (red/dashed) and ED (black/solid) results are shown.
 (b) The total hole occupancy of the compressed Ni (black/solid), 
 expanded Ni (red/dash-dot), and O (green/dashed) sites as obtained 
 from ED. 
 (c) As panel (b) but obtained within the HF treatment.  
 }
\end{figure}

Here we compare the results of the HFA with those obtained from 
exact diagonalization for the Ni$_2$O$_6$ cluster where periodic
boundary conditions  
have been assumed such that only the $\bk = 0$ momentum point is included. 
Fig. \ref{Fig:MF_vs_ED}a shows a comparison of the ground state energy
(electronic component only), identical 
to the results shown in Fig. 3(a) of the main text.  
In general the HF treatment overestimates the exact answer by $\sim 1$ eV, 
however the overall dependence on displacement is well reproduced.  
This general behavior also holds for the total hole occupancies, shown in 
Fig. \ref{Fig:MF_vs_ED}b and \ref{Fig:MF_vs_ED}c for the exact and 
HF solutions, respectively. This comparison gives us confidence that the 
HF method can describe the physics of the nickelates for our choice in 
parameters. 

